\pdfoutput=1
\documentclass[journal,final,letterpaper,oneside,onecolumn,11pt]{IEEEtran}
\usepackage{latexsym,url}
\usepackage{amsmath,amssymb,graphicx}% ,makeidx}
\usepackage{citesort}
\usepackage{epsf}

\long\def\comment#1{}

\newfont{\bbb}{msbm10 scaled 700}

\newfont{\bb}{msbm10 scaled 1100}

\newcommand{\uv}{{\bf u}}
\newcommand{\wv}{{\bf w}}
\newcommand{\vv}{{\bf v}}
\newcommand{\xv}{{\bf x}}
\newcommand{\yv}{{\bf y}}

\usepackage{graphics} % for pdf, bitmapped graphics files
\usepackage{epsfig} % for postscript graphics files
\usepackage{amsmath} % assumes amsmath package installed
\usepackage{amssymb}  % assumes amsmath package installed
\usepackage{dsfont}

\begin{document}

\title{\LARGE \bf Transmitting an analog Gaussian source over a Gaussian wiretap channel under SNR mismatch }

\author{
 Makesh Pravin Wilson and Krishna Narayanan\\
Dept. of Electrical Engineering \\
Texas A\&M University \\
College Station, TX 77843 \\
{\tt makeshpravin@neo.tamu.edu,krn@ece.tamu.edu}
 \thanks{This work was supported by the National
Science Foundation under grant CCR-0729210.} }
%\author{ Makesh Pravin Wilson}

\maketitle
%%% ----------------------------------------------------------------------

%%% ----------------------------------------------------------------------

%%% ----------------------------------------------------------------------
%

%Write the introduction at the last

\begin{abstract}
In this work we study encoding/decoding schemes for the transmission
of a discrete time analog Gaussian source over a Gaussian wiretap
channel. The intended receiver is assumed to have a certain minimum
signal to noise ratio (SNR) and the eavesdropper is assumed to have
a strictly lower SNR compared to the intended receiver.  For a fixed
information leakage rate $(I_{\epsilon})$ to the eavesdropper, we
are interested in minimizing the distortion in source reconstruction
at the intended receiver, and we propose joint source channel coding
(JSCC) schemes for this setup.  For a fixed information leakage rate
$(I_{\epsilon})$ to the eavesdropper, we also show that the schemes
considered give a graceful degradation of distortion with SNR under
SNR mismatch, i.e., when the actual channel SNR is observed to be
different from the design SNR.

% The encoder is designed for a fixed channel signal
%to
% noise(SNR) ratio, but the actual channel SNR may be larger than the
% designed SNR. The eavesdropper is assumed to have a fixed channel SNR that is strictly smaller than
%the intended receiver.
% We propose joint source channel coding schemes that give a graceful degradation
% of distortion with SNR, for a fixed information leakage $(I_{\epsilon})$ to the eavesdropper.
\end{abstract}

\section{Introduction, system Model and Problem Statement}

\begin{figure}[htb]
\begin{center}
\includegraphics[scale=0.8,angle =90]{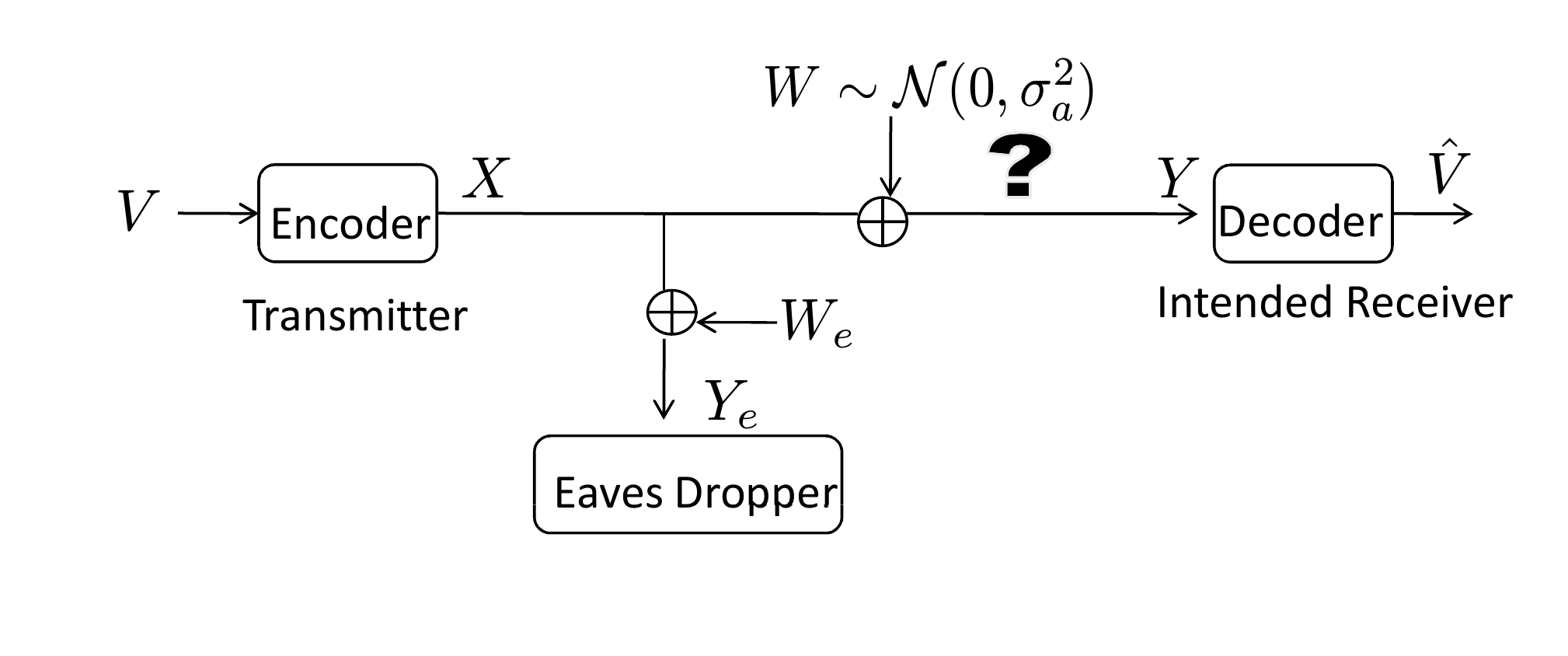}
%\center{\epsfxsize=5.0in \epsfbox[67 442 615 667]{problemmodel.eps}}
\caption{Problem model for the Secrecy system. }
\label{fig:problemmodel}
\end{center}
\end{figure}
Let us consider the classical case of transmitting a Gaussian source
over a Gaussian channel having an input power constraint, and we are
interested  in  estimating the source at the intended receiver with
the minimum possible  distortion. In this case, both the uncoded
scheme \cite{goblick1965theoretical} of scaling the source to match
the input power constraint and  the separation based scheme of
quantization followed by channel coding are both optimal for a given
channel SNR. Moreover, there are an infinite family of schemes that
can be shown to be optimal for the above setup
\cite{bross2006superimposed}. Compared to the separation based
scheme,  the uncoded scheme provides a graceful degradation of
distortion with SNR, though both schemes are optimal for a given
design $SNR_d$. In this work, we consider the same communication
setup but with an eavesdropper present, and we are interested in
minimizing the distortion for the intended receiver for a given
information leakage to the eavesdropper. For a fixed information
leakage $I_{\epsilon}$ to the eavesdropper, we propose a scheme
which achieves the minimal possible distortion for a given design
$SNR_d$ and also provides a graceful degradation of distortion with
SNR.

 The source
$V \sim \mathcal{N}(0,\sigma_v^2)$ is an $n$-length discrete time
real Gaussian  source and the channel is a discrete time real
Gaussian wiretap channel \cite{wyner1975wire}  as shown in Fig.
\ref{fig:problemmodel}. The source and channel bandwidths are
assumed to be matched and hence the input to the channel is a
$n$-length vector $\mathbf{x}$, which is also assumed to have an
average power constraint of $P$ expressed as $E[X^2] \leq P$.
Henceforth in this sequel, boldface is used to denote
$n$-dimensional vectors. The received signal at the eavesdropper
$\mathbf{y_e}$ can be expressed as
\begin{equation*} \mathbf{y_e} = \mathbf{x} + \mathbf{w_e}.
\end{equation*}
Here $\mathbf{w_e}$ is the n-length additive white Gaussian
noise(AWGN) vector having zero mean and noise variance of
$\sigma_e^2$. The intended receiver receives the n-dimensional
vector $\yv$ given by
\begin{equation*} \mathbf{y} = \mathbf{x} + \mathbf{w},
\end{equation*}
where the noise $\wv$  is  the n-length additive white Gaussian
noise vector having zero mean and noise variance of  $\sigma_a^2$.
The transmitter does not have an exact knowledge of $\sigma_a^2$ but
knows that $\sigma_a^2 \leq \sigma^2 $, where $\sigma^2$ is the
noise variance corresponding to some design SNR. The eavesdropper
channel is a degraded version of the main channel and is assumed to
have the lowest SNR i.e. $SNR_e  < SNR  < SNR_a $, where we define
$SNR_e := P/\sigma_e^2 $, $SNR_d := P/\sigma^2$ and $  SNR_a :=
P/\sigma_a^2$.  The receiver is assumed to have a perfect estimate
of $SNR_a$,  but the transmitter is assumed to be kept ignorant of
this information.

We refer to the information leaked to  the eavesdropper as the
information leakage rate and is expressed as $I_\epsilon :=
\frac{1}{n} I(\mathbf{V}; \mathbf{Y_e})$. The information leakage
rate is  the difference between the average source entropy
$\frac{1}{n} h(\mathbf{V})$ and the equivocation rate
$\frac{1}{n}h(\mathbf{V}|\mathbf{Y_e})$ defined in
\cite{wyner1975wire} and \cite{leung1978gaussian}. It is fairly
common to use  equivocation rate in literature, but since we have a
continuous source, the equivocation rate expressed as differential
entropy may not be always positive. Therefore, we choose the metric
as the mutual information between the source and the received signal
which is always $\geq 0$.  Notice that $I_\epsilon = 0$ corresponds
to perfect secrecy, and this implies that the eavesdropper gains no
information about the source.

 The intended receiver makes an
estimate $\mathbf{\hat{v}}$ of the analog source $\mathbf{v}$ from
the observed vector $ \mathbf{y}$. The distortion in estimating $V$
can be expressed as $D(SNR_a) = E[(V - \hat{V})^2]$, where $SNR_a =
P/\sigma_a^2$.  In this work for a fixed information leakage rate
$I_\epsilon$, we are interested in schemes which are optimal for
$SNR_d$ and also  which provide a low source distortion for
 $SNR_a > SNR_d$. We are also interested in studying the graceful degradation in distortion with
$SNR_a$. This can be captured mathematically by requiring the
exponent of $D(SNR_a)$ to be $-1$, or precisely
\begin{equation*}\lim_{SNR_a \rightarrow \infty} \frac{\log
D(SNR_a)}{SNR_a} = -1.\end{equation*} The exponent of $-1$ is
chosen, as it is the lowest possible exponent achievable for the
matched bandwidth case \cite{goblick1965theoretical}, even in the
absence of an eavesdropper.

 Before introducing our proposed schemes, we would like to mention some prior work in this area.
There has been a considerable amount of work in studying the
graceful degradation of distortion with $SNR_a$ for both the
bandwidth matched case \cite{goblick1965theoretical}  and the
bandwidth mismatched case, and several joint source channel coding
schemes have been proposed \cite{mittal2002hybrid}. There also
exists a considerable amount of literature on physical layer
security.
 The wire-tap channel was first introduced and studied by Wyner in \cite{wyner1975wire}, and the Gaussian wire-tap
channel was studied by  Leung and Hellman in
\cite{leung1978gaussian}. In \cite{yamamoto1997rate}, Yamamoto
studied the Shannon cipher system from a rate distortion
perspective, where in one of the theorems it is shown that for a
wiretap channel with a fixed SNR, a separation based approach of
quantization followed by secrecy coding is optimal. However in this
work, in contrast to \cite{yamamoto1997rate} we are interested in
JSCC scheme for the Gaussian wiretap channel under the SNR mismatch
case. We are also interested in quantifying the graceful degradation
of distortion with $SNR_a$ under the SNR mismatch case.
% \cite{TieliuShamai07} is one such work that looks at it and also
% the references therein point to several other papers that study this problem.

\section{A separation based scheme and a simple scaling scheme }
  For the classical setup without the eavesdropper, both the uncoded
scheme and the separation based scheme can be observed to be optimal
for a given design $SNR_d$. However these schemes perform
differently in the presence of a SNR mismatch. In this section we
look at the performance of both the  separation based scheme and the
uncoded scheme in the presence  of an eavesdropper.

\begin{figure}[htbp]
\begin{center}
\includegraphics[scale=0.8,angle =0]{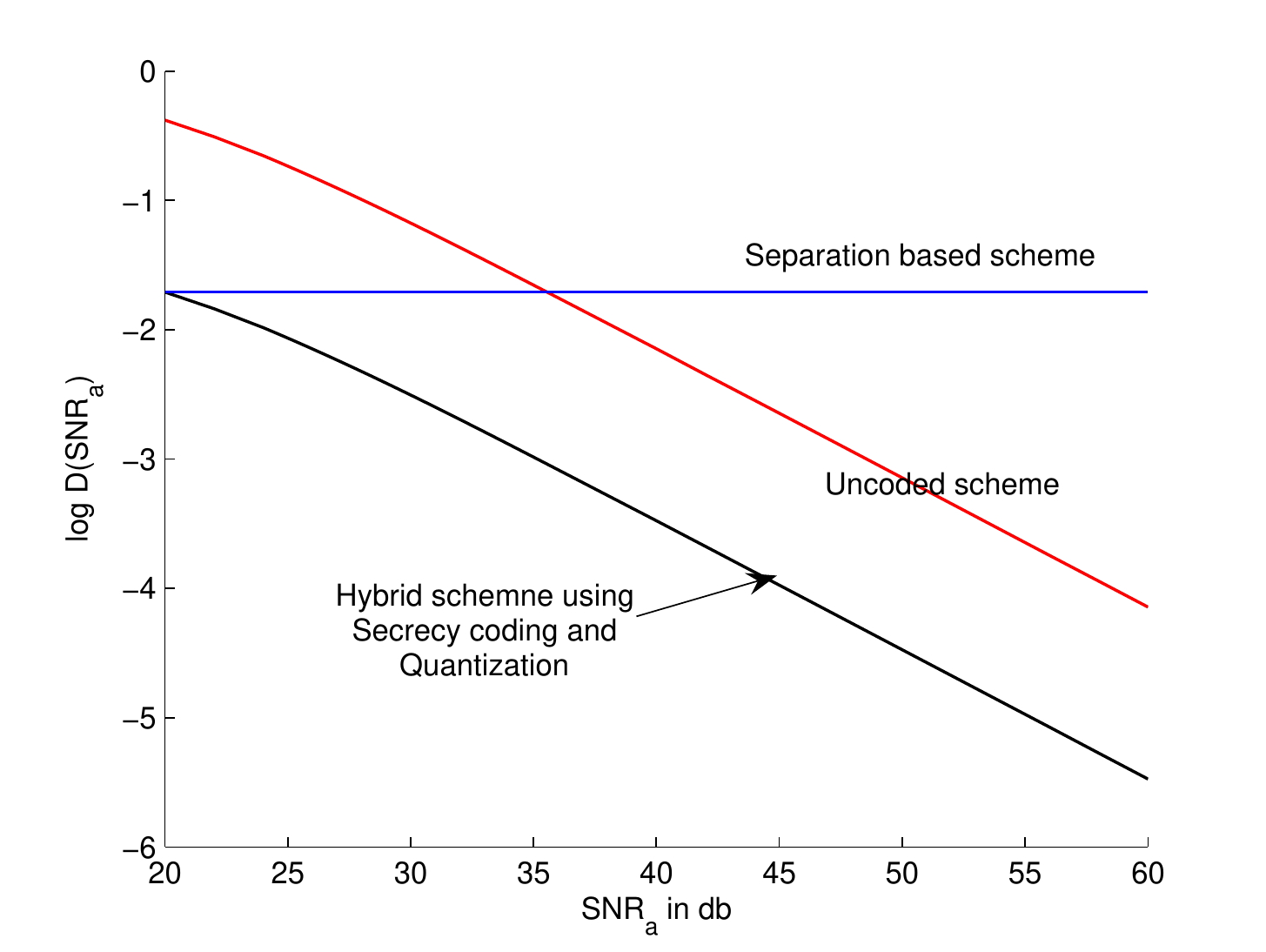}
%\center{\epsfxsize=4.5in \epsfbox[100 225 515 555]{schemesplot.eps}}
\caption{Distortion vs $SNR_a$ for different schemes for $I_\epsilon
= 0.01$. Here $P = 1$, $\sigma_e^2 = 1$ and $\sigma^2 = 0.01$($SNR_d
= 20db$). } \label{fig:schemesplot}
\end{center}
\end{figure}

\subsection{Separation based scheme}

Here we use a separation based scheme designed for a channel noise
variance of $\sigma^2$ and information leakage $I_{\epsilon} $. We
first design a vector quantizer of rate
\begin{equation*} R_v = C\left( \frac{P}{\sigma^2} \right) - C\left(
\frac{P}{\sigma_e^2} \right) + I_\epsilon,
\end{equation*} where  $C(x) := \frac{1}{2} \log(1 +
 x).$  We quantize $\vv$ to $\mathbf{v}_q$ using the designed vector quantizer of rate $R_v$.
$\mathbf{v}_q$ is then mapped to $\mathbf{v}_{sec}$ using a secrecy
code as in \cite{leung1978gaussian}. The encoded $\vv_{sec}$ is
transmitted over the channel. The eavesdropper obtains  a maximum
leakage rate of  $I_{\epsilon}$. The proof of this claim is
contained in the Appendix.  The intended receiver achieves a
distortion of $\sigma_v^2 2^{-2R_v}.$ Since we have quantized the
source to a fixed rate, the achievable distortion is constant for
all SNR's
 above the designed SNR and is given by,
\begin{equation*}
 D(SNR_a) = \sigma_v^2 2^{-2R_v} < \sigma_v^2.
\end{equation*}
For the given $I_\epsilon $, though we get some improvement in
distortion, the exponent is  $0$ and we do not have a graceful
degradation of distortion with $SNR_a$ as shown in
Fig.~\ref{fig:schemesplot}.

\subsection{A simple scaling scheme}
Let us first consider a scheme that is optimal for all channel SNRs
for the Gaussian wiretap channel in the absence of the eavesdropper.
This reduces to the classical problem of  point to point
communication over a Gaussian channel.  An optimal scheme is scaling
the analog source $\mathbf{v}$ by a constant $\kappa =
\sqrt{P/\sigma_v^2}$ to match the transmit power constraint
\cite{goblick1965theoretical}. The transmitted vector is hence given
by $\mathbf{x} = \kappa \mathbf{v}$.  The receiver has perfect
knowledge of $SNR_a$ and performs a minimum mean square estimate of
$\mathbf{v}$ from the observed $\mathbf{y}$. The distortion obtained
at the intended receiver is given by
\begin{equation*}D(SNR_a) = \frac{\sigma_v^2}{1 + SNR_a}.\end{equation*}
The distortion exponent can be seen to be $-1$ for this scheme and
also the obtained distortion is optimal for every given $SNR_a$.
 The information leakage rate is easily calculated to be
\begin{eqnarray}
\frac{1}{n} I(\mathbf{V}; \mathbf{Y_e}) = I(V; Y_e) = \frac{1}{2}
\log \left( 1 + \frac{\kappa^2\sigma_v^2}{\sigma_e^2}\right) =
\frac{1}{2} \log\left(1 + \frac{P}{\sigma_e^2}\right).\nonumber
\end{eqnarray}
From the above equation we can observe that the choice of $\kappa =
\sqrt{\frac{P}{\sigma_v^2}}$ results in a reasonably high
information leakage rate. However, we can reduce the value of
$\kappa$ to satisfy our eavesdropper information requirement of
$I_\epsilon$ as follows by choosing
\begin{equation*}
I_\epsilon =  \frac{1}{2} \log ( 1 + \frac{\kappa^2
\sigma_v^2}{\sigma_e^2})
\end{equation*} or
\begin{equation*}
\kappa^2 =  \frac{\sigma_e^2}{\sigma_v^2}(2^{2 I_\epsilon} - 1).
\end{equation*}

The intended receiver receives $ \mathbf{y} = \kappa \mathbf{v} +
\mathbf{w}$.
 Hence the distortion at the intended receiver is given by
 \begin{equation*}
 D(SNR_a) = \frac{\sigma_v^2}{1 + \frac{\kappa^2
 \sigma_v^2}{\sigma_a^2}}  = \frac{\sigma_v^2}{1 + \frac{\kappa^2
 \sigma_v^2 SNR_a}{P}}.
\end{equation*}

Hence for $SNR_a = SNR_d$, $D(SNR_d) > \sigma_v^2/(1 + SNR_d)$ as
$\kappa^2 \sigma_v^2 < P$. Though the  above equation shows that
 the distortion exponent is $-1$,  we have a
considerable loss in optimality at the intended receiver, as we have
not used the full power $P$ at the transmitter. This can be seen
from Fig.~\ref{fig:schemesplot} where for $SNR_a = 20$db , we see
that the uncoded scheme has a higher distortion performance compared
to the separation based scheme. However we can see from
Fig.~\ref{fig:schemesplot} that the uncoded scheme gives a graceful
degradation in distortion, unlike the separation based scheme. Also
if we need $I_\epsilon = 0$, then $\kappa$ must be chosen to be $0$
and hence $D(SNR_d) = \sigma_v^2,$ which is the worst attainable
distortion equivalent to simply estimating the source.

In the next section we show a scheme which is a combination of the
schemes mentioned above, that for $I_\epsilon \neq 0$ gives both the
optimal distortion for the designed SNR as well as a graceful
degradation in distortion for all SNR's above the designed SNR. In
the case of $I_\epsilon = 0$, or perfect secrecy, it is however not
clear if we can get a graceful degradation in distortion.

\section{ Hybrid scheme using secrecy coding and vector
quantization}

 In this scheme we quantize the source $\vv$ to get $\vv_q$ given as follows,
\begin{equation*}
\vv = \vv_q + \uv.
\end{equation*}
  The
 quantized digital part is encoded using a secrecy code and the quantization error $\uv$ is superimposed onto the secrecy code code and
 transmitted with some scaling. The transmitted vector $\xv$ is given
 by \begin{equation*} \xv = \vv_{sec} + \kappa \uv.  \end{equation*}
 Here $\vv_{sec}$ consists of a digital part that uses a secrecy code.

 The digital part $\vv_{sec}$ uses a power of $\alpha P$ and the analog
 part a power of $(1 - \alpha) P$.

 The source $\vv$ is quantized to $\vv_q$ at a rate $R(\alpha)$ chosen as
\begin{equation*}
R(\alpha) = C\left( \frac{\alpha P}{(1-\alpha)P + \sigma^2} \right)
- C\left( \frac{\alpha P}{(1-\alpha)P + \sigma_e^2} \right)
\end{equation*}

 The encoder is the same encoder as in the Gaussian wire tap
 channel case \cite{wyner1975wire}  and \cite{leung1978gaussian}.
We have $\approx 2^{nR(\alpha)}$ bins and $\approx 2^{nC\left(
\frac{\alpha P}{(1-\alpha)P + \sigma_e^2} \right)}$ codewords in
each bin. Hence the transmitted vector $\vv_{sec}$ has a rate of
$C\left( \frac{\alpha P}{(1-\alpha)P + \sigma^2} \right).$ The
intended receiver can decode the digital part and cancel $\vv_{sec}$
from $\yv$. It then forms a minimum mean square estimate (MMSE) of
$\uv$. Hence the obtained distortion can be expressed as
\begin{equation*}
D = \left.\frac{\sigma_v^2 2^{-2R(\alpha)}}{1 + \frac{(1- \alpha)
P}{\sigma_a^2}}\right.
\end{equation*}

\begin{figure}[htbp]
\begin{center}
\includegraphics[scale=0.8,angle =0]{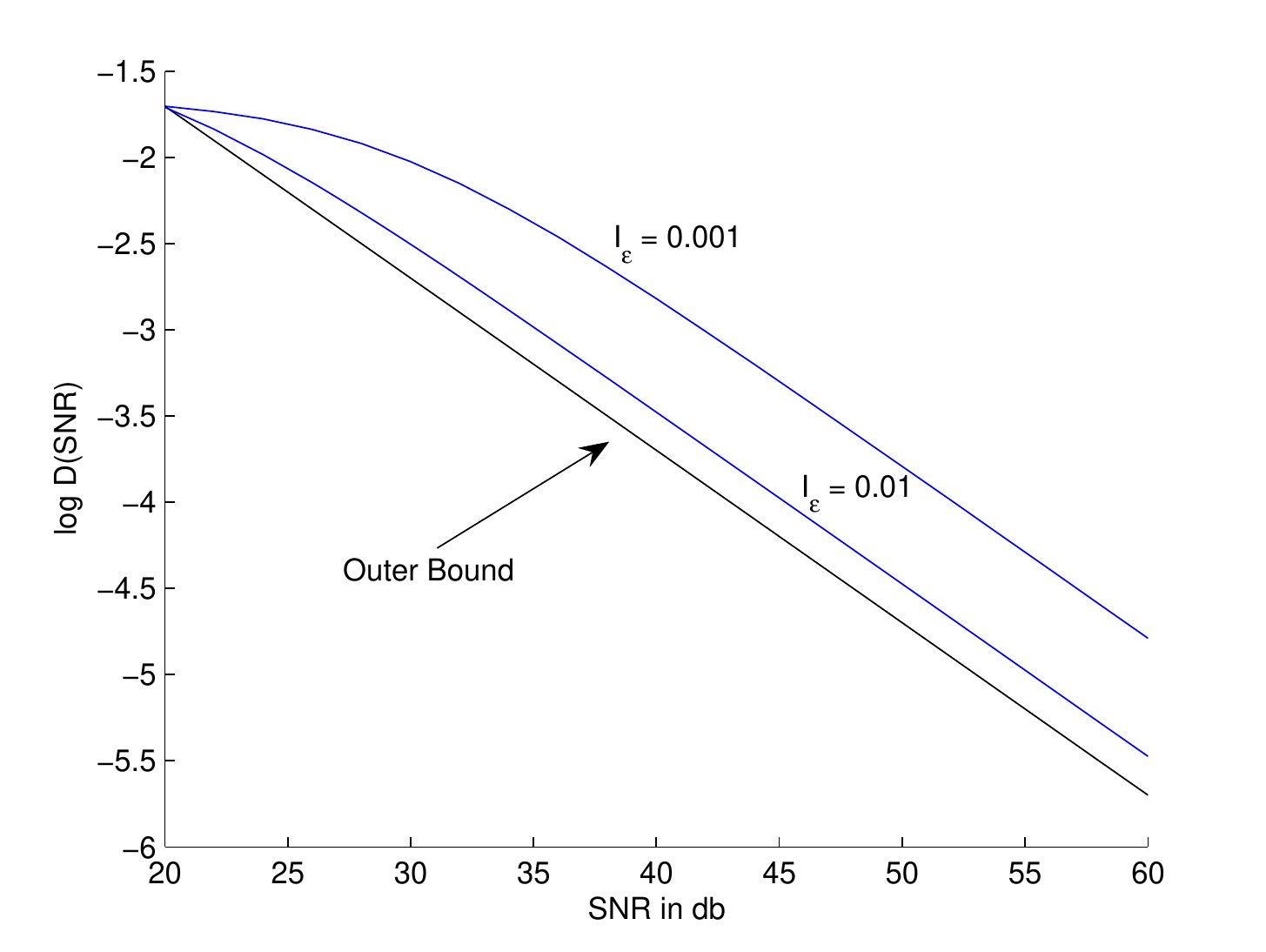}
%\center{\epsfxsize=4.5in \epsfbox[109 229 491 540]{distplot.eps}}
\caption{Distortion vs SNR for different $I_\epsilon$.  Here $P =
1$, $\sigma_e^2 = 1$ and $\sigma^2 = 0.01$($SNR_d = 20db$). }
\label{fig:distplot}
\end{center}
\end{figure}

The eavesdropper obtains $\yv_e$ and we are interested in
characterizing the information leakage rate $I_{\epsilon}$.
$I_{\epsilon}$ can be bounded as follows.
\begin{eqnarray}
I_{\epsilon} = \frac{1}{n} I(\mathbf{V}; \mathbf{Y_e})
&\stackrel{(a)}=& \frac{1}{n}
I(\mathbf{V_q}, \mathbf{\kappa U}; \mathbf{Y_e}) \\
         &\stackrel{(b)}=& \frac{1}{n}
I(\mathbf{V_q}; \mathbf{Y_e}) + \frac{1}{n}
I( \mathbf{\kappa U}; \mathbf{Y_e} | \mathbf{V_q}) \\
&\stackrel{(c)}=&  \frac{1}{n}
I( \mathbf{\kappa U}; \mathbf{Y_e} | \mathbf{V_q}) \\
&=&  \frac{1}{n}
h( \mathbf{\kappa U}| \mathbf{V_q}) -  \frac{1}{n}h( \mathbf{\kappa U}| \mathbf{Y_e} , \mathbf{V_q}) \\
&\stackrel{(d)}\leq&  \frac{1}{n}
h( \mathbf{\kappa U}) -  \frac{1}{n}h( \mathbf{\kappa U}| \mathbf{Y_e} , \mathbf{V_q},\mathbf{V_{sec}}) \\
&\stackrel{(e)}=&  \frac{1}{n}
h( \mathbf{\kappa U}) -  \frac{1}{n}h( \mathbf{\kappa U} - \beta(\kappa \mathbf{U} + \mathbf{W_e})| \mathbf{Y_e} , \mathbf{V_q},\mathbf{V_{sec}}) \\
&\stackrel{(f)}=&  \frac{1}{n}
h( \mathbf{\kappa U}) -  \frac{1}{n}h\left( (1-\beta)\kappa \mathbf{U} -\beta \mathbf{W_e} \right) \\
&\stackrel{(g)}=& \frac{1}{2} \log \left( 1 +
\frac{(1-\alpha)P}{\sigma_e^2}\right)
\end{eqnarray}

Here $(a)$ follows because of the Markov chain $V\rightarrow
(V_q,\kappa U) \rightarrow X$. $(b)$ follows from the chain rule of
mutual information. $(c)$ is obtained by the choice of our coding
scheme for $\mathbf{V_q}$, which is designed for perfect secrecy
from the eavesdropper. Hence $H(\mathbf{V_q}| \mathbf{Y_e}) =
H(\mathbf{V_q})$ or $I(\mathbf{V_q};\mathbf{Y_e}) = 0.$ We obtain
the first term in $(d)$ since $\mathbf{V_q}$ is independent of
$\mathbf{U}$, and the second term follows because conditioning
reduces entropy. In $(e)$, $\beta$ is chosen as $\beta =
\frac{(1-\alpha)P}{(1-\alpha)P + \sigma_e^2}$. $(f)$ follows because
$(1-\beta)\kappa \mathbf{U} -\beta \mathbf{W_e} $ is orthogonal to
$\mathbf{Y_e} , \mathbf{V_q}$ and $\mathbf{V_{sec}}$ and  finally
$(g)$ follows as all the terms are Gaussian. The distortion as a
function of $SNR_a$ is plotted in Fig.~\ref{fig:schemesplot} for
$I_\epsilon = 0.01$, which shows that the hybrid scheme performs
better than both the uncoded and the separation based scheme.
Fig.~\ref{fig:distplot} shows the performance of the hybrid scheme
for different values of $I_\epsilon$ and this  shows that the
distortion exponent is $-1$ for $I_\epsilon > 0$ . Also the
distortion that can be achieved at the eavesdropper can be lower
bounded by $\sigma_v^2 2^{-2I_\epsilon}.$ This can be seen to be
reasonable large when $I_\epsilon$ is small. Hence the eavesdropper
gets only a poor estimate of the source.

 A trivial outer bound for the problem can be obtained by
assuming that the transmitter has knowledge of $SNR_a$.
\cite{yamamoto1997rate} considers the rate distortion problem for
the Shannon cipher system. It can be  seen from \cite[Theorem 1 with
$R_k = 0$]{yamamoto1997rate}, that the Shannon cipher system reduces
to the wire-tap channel setup and the optimal distortion  can be
achieved by separate source coding followed by secrecy coding.
% The
%Shannon sipher system with \cite[$R_k = 0$]{yamamoto1997rate}
%reduces to the wire-tap channel setup. We evaluate the distortion as
%follows.
 We first quantize the source $\mathbf{v}$ to $\mathbf{v_q}$
at a rate $R$. For a maximal leakage rate of $I_\epsilon$, the
maximum value of $R = C\left(\frac{P}{\sigma_a^2} \right) -
C\left(\frac{P}{\sigma_e^2} \right) + I_\epsilon. $ This can be
obtained by following the steps outlined in the Appendix. Hence we
can achieve a distortion  $D = \sigma_v^2 2^{-2R}$ corresponding to
the above $R$. Fig.~\ref{fig:distplot} shows the outer bound and we
see that the achievable scheme has a constant gap from the outer
bound.

\section{Conclusion}

In this work we considered joint source channel coding schemes for
transmitting an analog Gaussian source over a Gaussian wiretap
channel. We analyzed the performance of a few schemes under SNR
mismatch. We showed that for a fixed information leakage to the
eavesdropper $I_\epsilon$, we can be optimal for a design SNR and
also can obtain a graceful degradation of distortion with SNR.  A
problem of interest is, wether it is still possible to  get this
graceful degradation when we enforce perfect secrecy at the
eavesdropper ($I_\epsilon = 0$)? Another possible future work is the
design of schemes for the source-channel bandwidth mismatch
scenario.

\appendix

 In this section we show that the information leakage rate is $I_\epsilon$ for
 $R_v = C\left(\frac{P}{\sigma^2}\right) -  C\left(\frac{P}{\sigma_e^2}\right) + I_\epsilon$.
Applying \cite[Theorem 1, (2), (3) and (17)]{leung1978gaussian} to
our problem setup we obtain the following set of equations.
\begin{equation*}
R_v = \frac{H(\mathbf{V_q})}{n}
\end{equation*}
%\begin{equation*}
%\Delta = \frac{H(\mathbf{V_q}| \mathbf{Y_e})}{H(\mathbf{V_q})}
%\end{equation*}
%\begin{equation*}
%R \leq C\left(\frac{P}{\sigma^2}\right)
%\end{equation*}
\begin{equation*}
R_v \frac{H(\mathbf{V_q}| \mathbf{Y_e})}{H(\mathbf{V_q})} \leq
C\left(\frac{P}{\sigma^2}\right) -
C\left(\frac{P}{\sigma_e^2}\right)
\end{equation*}

The left hand side term can be  simplified as follows,
\begin{eqnarray}
R_v \frac{H(\mathbf{V_q}| \mathbf{Y_e})}{H(\mathbf{V_q})}
&=& R_v
\frac{H(\mathbf{V_q}| \mathbf{Y_e}) - H(\mathbf{V_q}) +
H(\mathbf{V_q}) }{H(\mathbf{V_q})}\nonumber\\
&\stackrel{(a)}=& R_v \frac{H(\mathbf{V_q}) - n I_\epsilon}{H(\mathbf{V_q})}\nonumber\\
&=& R_v - I_\epsilon. \nonumber
\end{eqnarray}

In $(a)$ we used the definition of $I_\epsilon$. Thus we can bound
$R_v$ as,
\begin{equation*} R_v \leq C\left(\frac{P}{\sigma^2}\right) -
C\left(\frac{P}{\sigma_e^2}\right) + I_\epsilon.
\end{equation*}

Hence the maximal rate $R_v$ for the vector quantizer can be
obtained by choosing $R_v = C\left(\frac{P}{\sigma^2}\right) -
C\left(\frac{P}{\sigma_e^2}\right) + I_\epsilon.$

%\bibliographystyle{IEEEtran}
%\bibliography{IEEEabrv,bibfile}

\end{document}